\documentclass [12pt]{article}

\begin{document}
\begin{center}
\textbf{Epitaxial Growth of Thin Films -- a Statistical Mechanical Model}
\end{center}

\begin{center}
Anita Mehta
\end{center}

\begin{center}
{\it S N Bose National Centre for Basic Sciences,}
\end{center}

\begin{center}
{\it Block JD, Sector III, Salt Lake,}
\end{center}

\begin{center}
{\it Calcutta 700 098, India}
\end{center}

\begin{center}
and
\end{center}

\begin{center}
R. A. Cowley
\end{center}

\begin{center}
{\it Oxford Physics, Clarendon Laboratory,}
\end{center}

\begin{center}
{\it Parks Road}
\end{center}

\begin{center}
{\it Oxford OX1 3PU, U.K.}
\end{center}

{\bf Abstract}

\vskip 1cm

A theoretical framework is developed to describe experiments on the
structure of epitaxial thin films, particularly niobium on sapphire. We
extend the hypothesis of dynamical scaling to apply to the structure of thin
films from its conventional application to simple surfaces. We then present
a phenomenological continuum theory that provides a good description of the
observed scattering and the measured exponents. Finally the results of
experiment and theory are compared.

\vskip 1cm


\newpage
Although there has been great interest in the theoretical modelling of
molecular beam epitaxy (MBE) growth [1], there have been relatively few
attempts to compare the detailed predictions of the models with experimental
results. Also, models developed from statistical mechanics have mostly
concentrated on the shape of the growing interface rather than on the
structure of the grown material, whereas many experiments provide
information about the whole film. Microscopic models of thin films have been
developed [2] and successfully describe the structures of very thin films
consisting of only a few atomic layers, but far less has been achieved in
understanding the detailed structure of thicker films, 100-10000 {\AA}, that
are of importance for many applications. In this paper, we present a
dynamical scaling hypothesis for the structure of thin films involving the
displacement of the atoms from their ideal positions. Based on this, we
compare the results of x-ray scattering studies of thin films [3,4] with the
predictions of a continuum model for MBE growth developed by Villain [5].

It is well known [6], that for x-ray scattering from a single
interface, the relative intensity of the specular reflection
yields information about the mean-square width of the interface,
$<h(\textbf{x},t)^{2}>$, in the direction perpendicular to the
interface, while the form of the non-specular scattering provides
information about the correlation function $<h(\textbf{x},t)
h(\textbf{x}+\textbf{r},t)>$,  where \textbf{x} and \textbf{r} are
two-dimensional position vectors in the plane of the interface and
the averages are taken over all positions \textbf{x} on the
interface. The scattering from thin films requires more
discussion [7]. For a wave vector transfer \textbf{Q} the scattering
from a monatomic film is given by:

\begin{equation}
\label{eq1}
I({\rm {\bf Q}}) \propto {\sum\limits_{j,k} {\exp (i[{\rm {\bf Q}}_{\parallel}}
}.({\rm {\bf r}}_{{\rm {\bf j}}} - {\rm {\bf r}}_{{\rm {\bf k}}} ) + Q_{z}
(z_{j} - z_{k} )])
\end{equation}

\noindent
where (\textbf{r}$_{j}$, $z_{j})$ is the three-dimensional position of the
atom $j$, the summations are over all the atoms in the film and the growth
direction is parallel to the z-axis. We shall consider the scattering near
the Bragg reflection, with \textbf{Q = G} parallel to the growth direction
of the film so that $Q_\parallel = q$ and $Q_{z} = G+q_{z}$, with both
$q$ and $q_{z}$ being small compared with the reciprocal lattice
vector $G$. The positions of the atoms in the film are written in terms of the
atomic positions of the crystallographically perfect structure
(\textbf{R}$_{j}$, $Z_{j})$; the displacements from these positions for the
real structure are written as $z_{j} = Z_{j}$ + $u_{j}$. Eq. (1) can then be
rewritten as:

\begin{equation}
\label{eq2}
I({\rm {\bf Q}}) = {\sum\limits_{j,k} {\exp (i[{\rm {\bf q}}.({\rm {\bf
R}}_{{\rm {\bf j}}} - {\rm {\bf R}}_{{\rm {\bf k}}} ) + q_{z} (Z_{j} - Z_{k}
) + G(u_{j} - u_{k} )]) >}}
\end{equation}

\noindent
where it has been assumed that \textbf{q} and the displacements $u_{j}$ are
sufficiently small that terms of second order in these variables can be
neglected. The sum over the $z$-components is from the flat interface with the
substrate $Z = 0$, to the height of the film $Z_{j} = d + h_{j}$, where $d$ is
the average thickness of the film and the $h_{j}$ are the fluctuations in
the thickness whose average is zero. The Bragg-like component of the
scattering $I_{B}$ occurs for $\vert $\textbf{q}$\vert  = 0$; if the
distribution of the displacements and of the thickness fluctuations is
Gaussian, $I_{B}$ is given by:

\begin{equation}
\label{eq3}
I_{B} ({\rm {\bf Q}}) = D(q_{z} )\delta ({\rm {\bf q}})\exp ( - G^{2} <
u^{2} > ),
\end{equation}

\noindent
where $D(q_{z})$ is the Debye-Waller-like term arising from the fluctuations
in the thickness of the film, given in turn by:

\[
D(q_{z} ) = (1 - 2\cos (q_{z} d)\exp ( - q_{z} ^{2} < h^{2} > / 2) + \exp (
- q_{z} ^{2} < h^{2} > )) / q_{z} ^{2}
\]

This equation can then be further approximated to give:

\begin{equation}
\label{eq4}
I_{B} ({\rm {\bf Q}}) = 4q_{z} ^{ - 2}\sin ^{2}(q_{z} d / 2)\delta ({\rm
{\bf q}})\exp ( - G^{2} < u^{2} > - q_{z} ^{2} < h^{2} > )
\end{equation}

The leading term in the diffuse or non-Bragg-like scattering is obtained by
expanding the exponential term in Eq. 2 and assuming that the displacements
from the ideal positions and the fluctuations in the heights are
sufficiently small that only the leading terms need to be considered. The
expression for $q_{z} = 0$ is then:

\begin{equation}
\label{eq5}
I_{D} ({\rm {\bf Q}}) = {\sum\limits_{j,k} { < (d^{2}G^{2}\overline {u_{j}}
\overline {u_{k}}  + h_{j} h_{k} )\exp}}  (i{\rm {\bf q}}.({\rm {\bf
R}}_{{\rm {\bf j}}} - {\rm {\bf R}}_{{\rm {\bf k}}} )) >
\end{equation}

The displacements are averaged through the thickness of the film and are
written $\overline {u_{j}}  $, so that the summations over $j$ and $k$ are now
over all the lattice positions in the plane of the film.

Equations 3 and 4 generalise to a thin film the expressions for
the scattering from a single interface. The thin film has a
conformal structure if the fluctuations of $\overline {u} $ are
proportional to those of $h$. This corresponds to the case when
large fluctuations in the displacements are correlated with large
fluctuations in the film thickness. The two terms in the
scattering can then be combined into a single term. This
assumption of conformality enables us to propose an
\textit{extended dynamical scaling hypothesis for thin films.} The
scattering is then solely a function of the fluctuations in the
thickness/height instead of the two variables of the height and the
atomic displacements. Equations 3 and 4 form the basis of the
scaling hypothesis for the scattering from thin films.

With this assumption we now use conventional scaling relations for
the thickness of the film (which we have argued is conformal with the
atomic displacements) and their relationship to the scattered
intensity, to define the critical exponents $\alpha $ and $\beta $:

\begin{equation}
\label{eq6} u^{2}(t) = < h({\rm {\bf x}},t)^{2} > \approx
t^{2\beta} ,t \ll L^{z}, t, L \rightarrow \infty
\end{equation}

\noindent where $L$ is the size of the film and $z = \alpha /\beta$
 is the dynamical critical exponent,

\begin{equation}
\label{eq7}< h({\rm {\bf x}},t)h({\rm {\bf x}} + {\rm {\bf r}},t) > \approx {\left|
{{\rm {\bf r}}} \right|}^{2\alpha} ,{\left| {{\rm {\bf r}}}
\right|} \to \infty
\end{equation}

\noindent where the equalities above are valid
asymptotically, and the scattered intensity is
the Fourier transform of Eq. (7):

\[
 I({\rm {\bf q}}) = < h({\rm {\bf q}},t)h( - {\rm {\bf
q}},t) > 
\]

A continuum model of a surface was proposed by Villain [5] based
on the Edwards-Wilkinson (EW) equation [8] with the addition of a
surface diffusion term. The model leads to the differential
equation:

\begin{equation}
\label{eq9}
\partial h({\rm {\bf x}},t) / \partial t = - B\nabla ^{4}h({\rm {\bf x}},t)
+ A\nabla ^{2}h({\rm {\bf x}},t) + \eta ({\rm {\bf x}},t)
\end{equation}

The first term on the RHS models the surface diffusion, the second
is the normal diffusive term and $\eta $(\textbf{x},t) is the
 noise that is assumed as usual to have a Gaussian
distribution. When the surface diffusion term is
neglected, the equation reduces to the EW equation, which has been used as
an approximation to the KPZ [9] equation to model the growth of multilayers
[10]. We are unaware however of any comparison of the predictions of Eq. 8
with experimental data and in particular of the justification of the
presence of the different terms therein in the context of a specific
experiment.

In his important work on the theory of MBE growth, Villain [5]
pointed out the physical relevance of the terms in Eq. (8). In
the presence of desorption, the rate of change of $h(\textbf{x},t)$
can be written as the difference between the chemical potential
$\mu $ at a point $(\textbf{x},t)$ on the surface and the average
chemical potential $\mu _{c}$ of the atoms in the molecular beam:
arguing further that the consequent growth cannot depend on the
orientation of the plane leads to:

\begin{equation}
\label{eq10} {\frac{{\partial h({\rm {\bf x}},t)}}{{\partial t}}}
\propto (\mu ({\rm {\bf x}},t) - \mu _{c} ) = B\nabla ^{2}h({\rm
{\bf x}},t)
\end{equation}

A further source of the $\nabla ^{2}$h(\textbf{x},$t)$ term has
been argued to be the downward funnelling of atoms to the lowest
`valley' available [11] to them.

Turning to the surface diffusion term $\nabla
^{4}$h(\textbf{x},$t)$, the rate of change of $h$ is argued [4] to
obey a continuity equation with respect to a current density
\textbf{j,} which is itself the gradient of the chemical potential
$\nabla ^{2}$h(\textbf{x},$t)$:

\begin{equation}
\label{eq11} {\frac{{\partial h({\rm {\bf x}},t)}}{{\partial t}}}
= - \nabla .{\rm {\bf j}}({\rm {\bf x}},t) = - A\nabla ^{2}\mu
({\rm {\bf x}},t) = - A\nabla ^{4}h({\rm {\bf x}},t)
\end{equation}

This term is characteristic of MBE growth, and is typically known
as the noiseless Wolf-Villain equation [12,13]. It is a crucial
ingredient in any serious model of MBE growth. Finally we remark
that Eq. (8) is applicable to the motion of atoms on stepped
surfaces [5]: additionally, downward funnelling, finite desorption
and surface diffusion are valid physical processes and hence Eq.
(8) is a reasonable model for describing the experimental
conditions for MBE growth.

Equation (8) is linear and can be solved [5] exactly. If
the system is also isotropic, the solution depends on $q=\vert
\textbf{q}\vert $:

\begin{equation}
\label{eq12}
\begin{array}{l}
 {\left\langle {h(q,t)h( - q,t)} \right\rangle}  = {\left\langle {\eta
(x,t)\eta (x',t')} \right\rangle} \{1 - \exp [ - 2(Aq^{2} +
Bq^{4})t]\} /
(Aq^{2} + Bq^{4}), \\
 \\
 {\left\langle {h(q,t)h( - q,t)} \right\rangle}  = C_{0} [1 - \exp ( -
2\gamma (q)t)] / \gamma (q) \\
 \end{array}
\end{equation}

\noindent where $C_{0}$ is a constant depending on the amplitude
of the noise term and $\gamma (q)=Aq^{2}+Bq^{4}$. If we assume
that the film thickness $d$ is proportional to time, as is assumed
in theories of kinetic roughening [1], the diffuse scattering is
given by:

\begin{equation}
\label{eq13} I_{D} (q) = C_{1} [1 - \exp ( - 2C_{2} \gamma (q)d)] /
\gamma (q)
\end{equation}

\noindent where this expression follows from our dynamical
scaling hypothesis for thin films, and C$_{1}$ and C$_{2}$ are
constants.

Figure 1 shows the x-ray scattering observed [3] from one of the thin films of
niobium grown on sapphire single crystal substrates. The thickness
of all the films is larger than the critical thickness and so the
lattice parameters of the films are relaxing towards the bulk
niobium lattice parameter. The solid line shows the result
of a least-squares  fit of the parameters $C_{1}$,  $C_{2}$, $A$ and $B$ 
 in Eq. (12) to
the data and we see immediately that Eq. (12) provides a very
satisfactory description of the diffuse scattering. In particular
for large $q$, the intensity is inversely proportional to $\gamma
(q)$ and so describes the $q^{ - 4}$ wings of the scattering [3]
very well, while at small $q$ the theory has a finite value of the
diffuse scattering as found in the experiments. Similar good agreement
was obtained for the other layers and for the scattering around the
(110) Bragg reflection.

The full width at half maximum (FWHM) of the diffuse scattering in $q$ was
 found [3] to vary with
the layer thickness $d$ as $d^{ - 0.51\pm 0.05}$. Assuming as usual that
thickness is proportional to time, this suggests that the FWHM varies as
$t^{ - 1 / 2}$. The correlation length of the model is defined [1]
by $\xi (t) \sim  t^{1 / z}$ and the FWHM is proportional to
$1/\xi $. The value of the exponent then implies that the
behaviour of the FWHM is dominated by the diffusive, $\nabla
^{2}h(x,t)$, term in eq. (8). Since the line-shape for large $q$
requires the quartic term, these two results strongly suggest that
{\it both} the quadratic and quartic terms in Eq. (8) are needed to
explain the experimental results.

The experiments also measured the mean square fluctuation of the
structure as a function of the thickness and found that $u^{2}(d)
 \sim  d^{0.68\pm 0.05}$, implying that
  the temporal roughening
exponent is $\beta  = 0.34\pm 0.03$.This value of $\beta $
 is the value that is obtained if $A =
0$ for the solution of Eq. (8) for a one-dimensional surface; it is
the value corresponding to the Wolf-Villain model i.e., with only the
quartic term in Eq. (8). We therefore note that the behaviour of both $u(d)$
and the FWHM are in agreement with respectively the $A = 0$ (quartic term only)
 and $B 
=  0$  solutions (quadratic
term only) of Eq. (8) for a one-dimensional surface. Normally we would expect that a two-dimensional system would
exhibit logarithmic roughening [1], and we shall discuss below
possible reasons for a dimensionality shift.

We note that the spatial roughening exponent for the
one-dimensional Wolf-Villain equation, (10), is $\alpha =1.5$
which is an indication of anomalous scaling behaviour [1]. Since
the $\beta $ exponent deduced from the measurements of $u(d)$ is in
good accord with the Wolf-Villain model we consider that at least
in some regime the films should exhibit anomalous scaling.
 There are some indications of this in Fig. 4 of ref. [3]; the
scattering widths as a function of $q_{x}$, i.e. perpendicular to the growth
direction showed an interesting behaviour. For small $q_{x} $ (and
corresponding to a lengthscale larger than the film thickness $d$), they were
largely independent of $q_{x}$, but for larger $q_{x}$, they increased
approximately linearly with $q_{x}$. This was earlier [3] given the
interpretation that in the latter case, the displacements with these
wavevectors did not propagate through the film, while in the former case,
the scattering profile was characteristic of the whole thickness of the
film. Another way of seeing this, however, is to say that the distribution
of the atomic displacements $<u_{z}^{2}>$ depends on the length scale over
which it is examined. If the film is
conformal, it is conceivable that this inhomogeneity in the distribution of
atomic displacements
should be reflected in surface fluctuations, which would then exhibit the
same lengthscale dependence, leading in turn to
anomalous scaling.

Other experimental measurements of the spatial roughening exponent
in metallic materials have given values of $\alpha  = 0.79\pm
0.05$ [14], $0.65\pm 0.03$ [15] and $0.85\pm 0.05$ [16]. These
values are intermediate between the values of the exponents
predicted by the one-dimensional EW equation ($\nabla ^{2}h$), 
$1.5$, and that of the Wolf- Villain model ($\nabla ^{4}h$),  $0.5$,
 which
are the two limiting cases of our Eq.(8). Further investigations are needed
to determine whether the experimental systems need entirely different
theoretical models to describe them, or whether they could be treated as
intermediate cases that could be accommodated within the general framework
of our present work.

Apart from the values of the exponents, the fits to the
experimental results provide information about the values of the
parameters in Eq. (11). The measurements were made in such a way
that the scattered intensity could not be reliably scaled from one
sample to another and so the constants $A$ and $B$ were obtained by
adjusting the overall scale factor to give the appropriate total
scattered intensity for the thickness of the film. The results for
parameter $A$ are shown in Fig. 2. There is considerable scatter in
the results at least partly because the parameters in the least squares
fits are strongly coupled, and both $A$ and $B$ are consistently larger when deduced from the (110) data in comparison
with the (220) data. As
discussed in [3], some of the differences between the films may
arise because they were grown at different times and capped in
different ways. The difference between the results obtained from
scattering near the (110) and (220) Bragg reflections may arise
from a failure of the small displacement approximation. Also, the
small value of $A$ obtained for the film with a thickness of
$1500${\AA} may be because the correlation length is then so small
that it is difficult to distinguish between the Bragg component
and the diffuse scattering.

Figure 3 shows the thickness dependence of the $B$ parameter. This
term arises from the quartic term in the differential equation (8)
and so to compare it with the $A$ term we need to scale it by the
square of a length. In Fig. 3 we show $B/d^{2}$ because we found
empirically that this quantity was largely independent of $d$.

A full microscopic theory of the structure of thin films is not
yet available. The distortions from a perfect structure arise
presumably from the existence of threading dislocations as
explained, for example, by Kaganer et al [17]. Although their
theory can be used to describe thin films with very low
dislocation densities or for very thick films, it is the
intermediate region that is important for many applications.

In conclusion, we have proposed a scaling ansatz to describe the development of the
structural deformations in thin films and have applied it to describe the
structure of Nb thin films grown on sapphire substrates. We note that other
films, both semiconductors and metals give similar experimental results. The
scattering is well described by the profile proposed by Eqs. (8, 12). The
exponent measured for the spatial correlation length shows that it is
dominated by the diffusive interactions (quadratic term) in the
phenomenological differential equation (8), while the quartic term in Eq.
(8) is needed to explain the temporal roughening exponent. We believe we may
in this sense have provided a minimal model to explain various quantitative
features of the experiment in ref. [3].

 Normally we would expect that a two-dimensional system such as the
corresponding to the experiment in [3] would exhibit logarithmic roughening
[1], rather than the power-law roughening indicated by our measurement of
the temporal roughening exponent, $\beta $. Since the simplest theories
state that two-dimensional surfaces are at the critical dimension for which
the roughening shows logarithmic behaviour when the diffusive term is
present, the power law result is unexpected. We note that good agreement
with the observed exponent is obtained instead for the linear
one-dimensional theory presented here. 

The most likely explanation for this is that since
the measurements concerned were the result of synchrotron experiments,
there was a high degree of anisotropy in their resolution. In particular,
in the experiments [3], 
the experimental resolution element perpendicular to the
scattering plane was much larger than in the other directions; the integrated
data over the x-y plane could also have contained some highly anisotropic averaging.
 It could also be that there is in the experiment [3] a
dimensionality shift analogous to that found in the random field model [18]
due to the initial defects present in the substrate. Another remark concerns
the fact that the ratio $B/d^{2}$ is approximately constant (Fig. 3), while
$A$ decreases with increasing $d$, leading to the increasing dominance of $B$
 for
large film thicknesses. This could be because as the thickness of the film
increases, with its length always remaining constant, the approach to
asymptoticity in eq. (8) is slowed down, as a result of which the
higher-order nonlinear term (whose coefficient is $B$, in this case) appears
to win out. We hope that the relative success of our minimal model in
explaining quantitative observations on the sapphire-Nb film analysed in [3]
will lead to some of these issues being resolved, and to new theoretical
developments in the study of thin films and multilayers , for which the
thickness is in the important range between $100$ and $10000$ {\AA}.

\vskip 2cm

{\bf Acknowledgements}

AM is grateful for the hospitality of the Department of Physics in Oxford,
where she held an EPSRC Visiting Fellowship for a year, during the course of
which most of this work was carried out. We have benefited from helpful
discussions with Drs A. Babkevich, D. F. McMorrow and A. R. Wildes and from
financial support by EPSRC. We are very grateful to Prof. Metin Tolan for
very stimulating discussions and a critical reading of the manuscript.

\vskip 1cm

{\bf Figure captions}:

Fig.1:\\ 
X-ray scattering intensity from a transverse scan of the wave-vector
transfer through the (220) Bragg reflection from the Nb layer, 20nm thick.
The observed intensity is fitted to a Gaussian to represent the Bragg
reflection and the function proposed in the text for the diffuse scattering.

Fig. 2: \\
The parameter $A$ as a function of the layer thickness as deduced from
scattering near the (110) Bragg reflection (filled points) and the (220)
reflection (open points).

Fig. 3: \\
The parameter $B/d^{2}$ as a function of layer thickness. The data
were obtained from scattering near the (110) Bragg reflection (filled
points) and the (220) reflection (open points).

\vskip 2cm

{\bf References}:

1. J. Krug, Advances in Physics \textbf{46} 139 (1997)

2. S. Das Sarma and P. Tamborenea Phys. Rev. Lett. 66 325 (1991);S. Das
Sarma and S. V. Ghaisas, Phys. Rev. Lett. 69 3762 (1992)

3. A. R. Wildes et al., J. Phys: Condens. Matt. \textbf{10} L631 (1998)

4. P. M. Reiner et al., Phys. Rev. \textbf{45} 11426 (1992), G. Gutekunst et
al., Phil Mag. A \textbf{75} 1329, 1357 (1997) and P. F. Micelli et al.,
Physica \textbf{221} 230 (1996)

5. J. Villain, J. de Physique I \textbf{1} 19 (1991)

6. S. K. Sinha et al., Phys. Rev. B \textbf{38} 2297 (1988)

7. V. Holy, U. Pietsch, and T. Baumbach; \textit{High resolution x-ray}

\textit{scattering from crystalline thin films}, Springer Tracts in Modern Physics, Vol.149, Springer-Verlag (1999).

8. S. F. Edwards and D. R. Wilkinson, Proc. Roy. Soc. A \textbf{381} 17
(1982)

9. M. Kardar et al., Phys. Rev. Lett. \textbf{56} 889 (1986)

10. T. Salditt et al., Phys. Rev. Lett. \textbf{73} 2228 (1994)

11. D. E. Wolf and J. Villain, Europhys. Lett. \textbf{13} 389 (1990)

12. J. W. Evans, Phys. Rev. B \textbf{43} 3897 (1991)

13. G. L. Zhou and C. P. Flynn, J. Phys: Condens. Matt. \textbf{9} L671
(1997)

14.Y. L. He et al., Phys. Rev. Lett. \textbf{69} 3770 (1992)

15. X. Yan and T. Egami, Phys. Rev. B \textbf{47} 2362 (1993)

16. P. P. Swaddling et al., Phys. Rev. Lett. \textbf{73} 2232 (1994)

17. V. M. Kaganer et al., Phys. Rev. B \textbf{55} 1793 (1997)

18. Y. Imry and S. K. Ma Phys. Rev. Lett. \textbf{35} 1399 (1975)

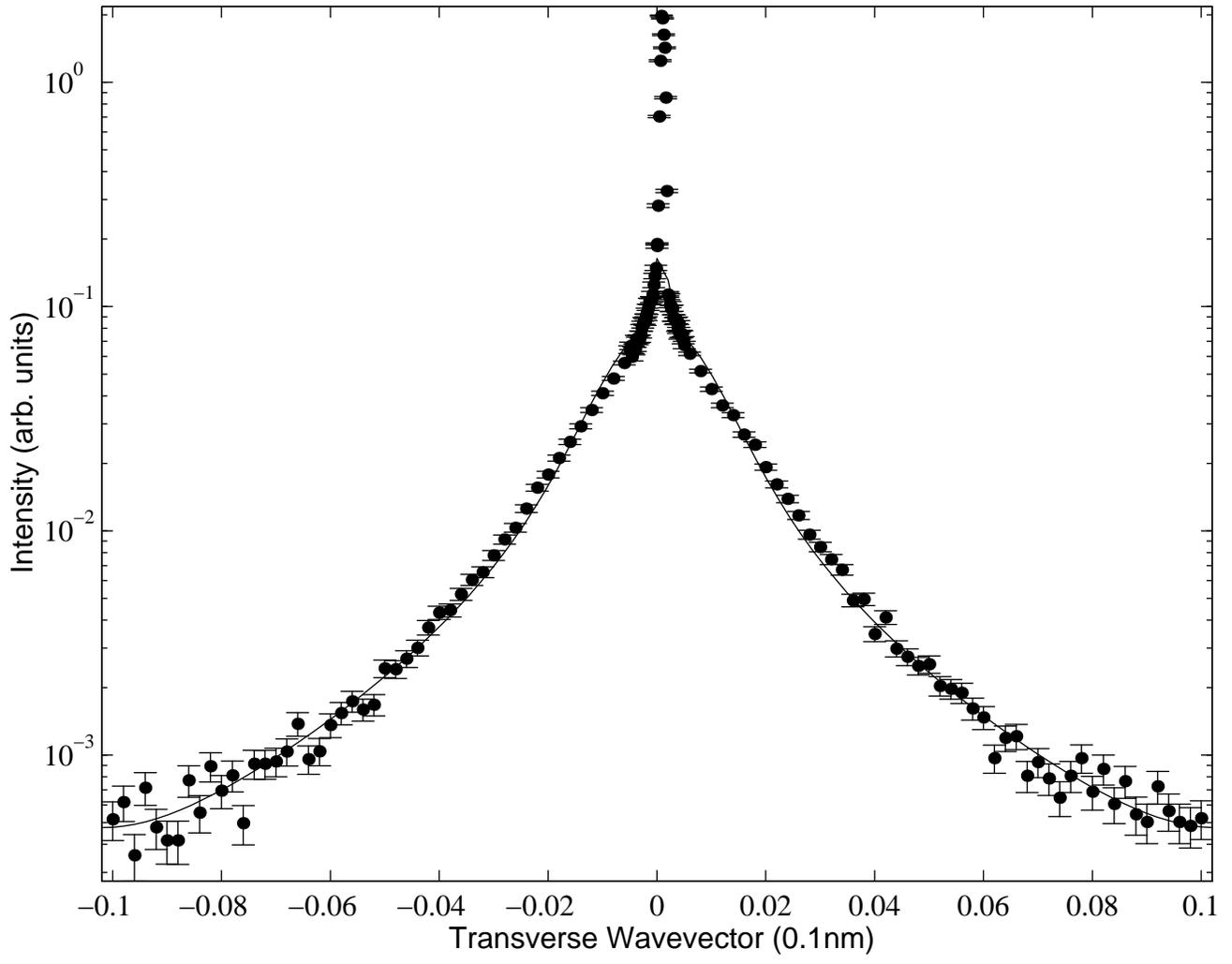
\end{document}